\documentstyle[aps]{revtex}
\newcommand{\ds}{\displaystyle}
\newcommand{\dsf}{\ds\frac}

\newcommand{\beq}{\begin{equation}}
\newcommand{\eeq}{\end{equation}}

\large
\begin{document}
\large
\begin{center}
\Large\bf
  STABILITY OF NONLINEAR STATIONARY WAVES IN COMPOSITE SUPERCONDUCTORS
\vskip 0.1cm
{\normalsize\bf N.A.\,Taylanov}\\
\vskip 0.1cm
{\large\em Theoretical Physics Department and
Institute of Applied Physics,\\
National University of Uzbekistan,\\
E-mail: taylanov@iaph.silk.org}
\end{center}
\begin{center}
\bf €bstract
\end{center}

\begin{center}
\mbox{\parbox{14cm}{\small
The thermomagnetic instability of the critical state in superconductors
is analysed with account of the dissipation and dispersion.
The possibility is demonstrated of the existance of a nonlinear shok
wave describing the final stage of the instability evolution in a
superconductor. The structures possess a finite-amplitude and propagate
at a constant velocity. The apperance of these structures is qualititively
described and the wave propagation velocity is estimated.
The problem of nonlinear wave stability with respect
to small thermal and electromagnetic perturbations. It is shown that only
damped perturbations correspond to space-limited solutions.
}}
\end{center}
\vskip 0.5cm
        Dissipative processes caused by Joulean heat-release in viscous
motion of Abrikosov vortices give rise to different regimes of wave
processes describing the final stage of evolution of thermomagnetic
and electromagnetic perturbations in the critical state of type II
superconductors.
A typical example of these regimes is the nonlineat stationary
thermomagnetic wave discovered in early theoretical investigations [1-4].
It was shown in [1], that depending on the conditions at the supercondutor
surface there may exict nonlinear stationary thermomagnetic waves of two
types $\vec E$ or $\vec H$. The problem of the stability of nonlinear waves with
respect to small thermal and electromagnetic perturbations in superconductors
has not yet been investigated well enough.
         The stability of nonlinear thermomagnetic waves with respect to
small thermal and electromagnetic perturbations in superconductors in the
critical state is the subject of this study. It is shown that only damped
perturbations correspond to space-limited solutions.
        The evolutions of the of the thermal $(T)$ and electromagnetic
$(\vec E, \vec H)$ perturbations is described by the nonlinear equation of the
thermal conductivity [1]
\beq
\nu\frac{dT}{dt}=\Delta [\kappa\Delta T]+\vec j\vec E,
\eeq
by Maxwel's equations
\beq
rot\vec E=-\dsf{1}{c}\dsf{d\vec H}{dt},
\eeq
\beq
rot{\vec H}=\dsf{4\pi}{c}\vec j
\eeq
and by the equation of the critical state
\beq
\vec j=\vec j_{c}(T,\vec H)+\vec j_{r}(\vec E).
\eeq
where $\nu=\nu(T)$ and $\kappa=\kappa(T)$ are the heat capacity and thermal
conductivity respectively; $\vec j_c$ is the critical current density and
$\vec j_r$ is the resistance current density.

        Let us consider a planar semi-infinite sample $(x>0)$ placed in an
external magnetic field $\vec H=(0, 0, H_{e})$ growing at a constant rate
$\dsf{d\vec H}{dt}=const$. According to Maxwel's equation (2), there is a vortex
electric field $\vec E=(0, E_e, 0)$ in the sample, directed parallel to the
current density $\vec j$: $\vec E\parallel \vec j$; where $H_e$ is the
amplitude of the external magnetic field and $E_e$ is the amplitude of the
external electric field.

          Let us use the Bean-London model of the critical state for the
dependence $j_{c}(T,H)$; $\dsf{dj_c}{dH}=0$ [5]. According to this model
the dependence $j_{c}(T)$ can be represented as
$j(T)=j_0[1-a(T-T_{0})]$: where $j_{0}$ is the equilibtium current density,
$a$ is the constant parameter [3] and $T_0$ is the temperature of the cooling
medium.

        At sufficiently high electric fields $(E>E_f)$ the
typical dependence $j_{r}(E)$ can be approximated by a piecewise-linear
function, $j_r\approx\sigma_f E$ where as at small fields $E<E_f$
the dependence $j_{r}(E)$ is, nonlinear. The nonlinearity is due to the
thermo-activation creep of the flow [6].

        For the automodelling solution of the form $\xi=x-vt$, describing
a travelling wave moving at a constant velocity $v$ along the $x$ axis,
the system of equations takes the following form
\beq
- \nu v[T-T_0]=\kappa\dsf{dT}{d\xi}-\frac{c^2}{4\pi v}E^2,
\eeq
\beq
\dsf{dE}{d\xi}=-\dsf{4\pi v}{c^2}j,
\eeq
\beq
E=\dsf{v}{c}H.
\eeq

       The thermal and electrodynamic boundary conditinos for equations
(5)-(7) are as follows:
\beq
\begin{array}{l}
T(\xi\rightarrow+\infty)=T_0, \dsf{dT}{d\xi}(\xi\rightarrow-\infty)=0,\\
\quad\\
E(\xi\rightarrow+\infty)=0,   E(\xi\rightarrow-\infty)=E_e,\\
\end{array}
\eeq

        Excluding the variables $T$, $H$ by using (5) and (7) with the
boundary conditions (8) we obtain a differential equation for the
distribution $E(\xi)$:
\beq
\dsf{d^2 E}{dz^2}+\beta(1+\tau)\dsf{dE}{dz}+\beta^2\tau E-
\dsf{E^2}{2E_\kappa}=0
\eeq

Here the following dimensionless parameters are introduced
$z=\dsf{\xi}{L}, \beta=\dsf{vt_\kappa}{L}$;
$L=\dsf{cH_e}{4\pi j_0}$ is the depth of the magnetic field penetration
into the superconductor, $\tau=\dsf{D_t}{D_m}$ is the parameter describing
the ratio of the thermal $D_{t}=\dsf{\kappa}{\nu}$ to the magnetic
$D_{m}=\dsf{c^2}{4\pi\sigma_{f}}$ diffusion coefficient;
$t_\kappa=\dsf{\nu L^2}{\kappa}$ is the time of thermal diffusion, and
$E_{\kappa}=\dsf{\kappa}{aL^2}$ is a constant.

        In the approximation of weak superconductor heating $(T-T_0)<<T_0$,
the heat capacity $\nu$ and the thermal conductivity $\kappa$
depend on temperature profile only slightly. Then at $\tau>>1$ we can
neglect the terms with dissipative effects in equation (9) and represent
its solutions as [7]
\beq
E(z)=\dsf{E_1}{2}\left[1-th\dsf{\beta\tau}{2} (z-z_0)\right].
\eeq

        Using the boundary condition $E(z\rightarrow-\infty)=E_e$
we readily determine the velocity $v_{E}$:
\beq
v_E=\dsf{L}{t_\kappa}\left[\dsf{E_e}{2\tau E_\kappa}\right]^{1/2},
\eeq
of a wave with amplitude $E_e$ and front width
\beq
\delta z=16\dsf{(1+\tau)}{\tau^{1/2}}\left[\dsf{E_\kappa}{E_e}\right]^{1/2}.
\eeq
Numerical estimations give $v_E=1\div 10^2 \dsf{sm}{sec}$ and
$\delta z=10^{-1}\div 10^{-2} $ for $\tau=1$.

         Expression (10) describes the profile of the thermomagnetic
shock wave travelling into the supercondutor.

        To investigate the stability of the nonlinear wave with respect
to small perturbations, let us represent the solution of the system
(1)-(4) as
\beq
T(z,t)={T}(z)+\delta T(z,t) \exp\left[\frac{\lambda t}{t_\kappa}\right],
E(z,t)={E}(z)+\delta E(z,t) \exp\left[\frac{\lambda t}{t_\kappa}\right],
\eeq

where ${T}(z)$ and ${E}(z)$ are the stationary solutions, and $\delta T,
\delta E$ are small perturbations. Substituting (13) into (1)-(4) and
assuming that $\delta T$ and $\delta E<<{T},{E}$ at $\tau>>1$, which
corresponds to a "slow' instability $\lambda<<\dsf{\nu v^2}{\kappa}$ [6]
we obtain the following equations for determining the eigenvalues
$\lambda$ :
\beq
[H-\Lambda]\Psi=0
H=\frac{d^2}{dy^2}+U(y),
U(y)=-1+\frac{2}{ch^2}+\Lambda thy.
\eeq
where
$\Psi(y)=\delta E(y) chy, y=\dsf{\beta\tau}{2} (z-z_0)$,
$\Lambda=\dsf{\lambda}{\beta^2}$.

   The stationary wave is stable if the eigenvaluesof the operator $H$
include negative $\Lambda<0$. As a boundary conditions for (14) cerves
the condition that the solutions $\Psi(y)$ are bounded at infinity.
The odd term in (14) describes the effect of the thermal mode on the
dynamics of electromagnetic perturbations in the superconductor.
Using a standard procedure like that in [8], we can readily find the
exact solution of equation (14) in the form
\beq
\Psi(y)=(1-thy)^{p+\frac{1}{2}}(1+thy)^{q+\frac{1}{2}}F\left(\alpha,\beta,
\gamma,\dsf{1-thy}{2}\right),
\eeq
Here $F$ is a hypergeometric function [9];
$\alpha=p+q+3$, $\beta=p+q$, $\gamma=2p+1$.
The constant coefficients $p$ and $q$ are determined fr®m quadratic
equations:

$p^2-\frac{\Lambda}{4}-\frac{1}{4}=0$,
\quad $q^2-p^2-\frac{\Lambda}{2}=0$

It can be seen that the spectrum of eigenfunctions of equation (14) is
continuous. Analysis of the asymptotic behaviour of solution (15) at
$y\rightarrow\pm\infty$ shows that $\delta E(y)$ is bounded only at
$\Lambda<0$. Indeed, one of partial solutions of equation (14), remaining
bounded at $y\rightarrow-\infty$ has the form:
\beq
\delta E(y)\approx  \exp(\omega_p y)F(p+q+2,p+q-1,2p+1,exp(-2y)); 1-2p=\omega_p,
\eeq

It is obvious that at $y\rightarrow+\infty$ $(exp(-2y)\rightarrow0)$
to the asymptotics (16) corresponds a wave exponentially decaying at
positive $\omega_{p}$ and growing as $exp(i\omega_{p}y)$ at negative
$\omega_{p}$. It may be assumed that the obtained solution exhibits a
regular asymptotic behaviour at $y\rightarrow+\infty$,i.¥. the function
$\delta E(y)$ is bounded only at those eigenvalues for which
$\Lambda<0$.

        Let us represent the second linear-independent solution of
equation (14) as:
\beq
\begin{array}{l}
F(p+q+2, p+q-1, 2p+1, s)= \dsf{ƒ(2p+1) ƒ(-2q)}{ƒ(p-q-1) ƒ(p-q+2)}\times\\
\times F(p+q+2, p+q-1, 2p+1, 1-s)+ \\
+\dsf{ƒ(2p+1) ƒ(2q)}{ƒ(p+q+2) ƒ(p+q+1)}
(1-s)^{-2q} F(p-q-1, p-q+2, -2q+1, 1-s),
\end{array}
\eeq

where ƒ is Euler's $\gamma$ function [9]. At $y\rightarrow-\infty$ the
following asymptotic representation is valid
\beq
\delta E(y)\approx exp(\omega_{q}y)$; $1-2q=\omega_q.
\eeq
It can be seen that the space-limited thermomagnetic perturbations
are damped both at $\omega_p>0$ and at $\omega_q>0$.

In conclusion, it should be noted that the description of the critical
state used in this study is based on the microscopic theory of Bardeen,
Cooper and Schriffer (BCS). We use the contunial approximation and
assume that the physical characteristics of the system change weakly
on a scale corresponding to the mean distance $d$ between the vortices.
The continuity condition has the form $L,\delta z>>d$. A nonlinear wave
can be exist in samples with thickness $2l$ satisfuing the condition
$\delta z, L << l$.

        On more restrition appears in view of the possible overheating
of the superconductor to above the critical temperature $T_c$, when
the system of equations (1)-(4) is not valid. Such a situation occurs
if the condition $j_{c}(T_0,E_e)>a(T_c-T_0)$ is satisfied, $T_c$ is the
critical temperature. In this case,
a zone of normal conductivity must appear immediately behind the wave
front. This zone stimulates additional energy dissipation in the
superconductor.

        The thermomagnetic wave $E$ can be observed experimentally
by initiating motion of a magnetic flux on the background of an
external magnetic field growing at a constant rate $\dot H_e$.
Then a vortex electric field whith amplitude $E_e$ exists in the
superconductor from the beginning.

\end{document}